\documentstyle[11pt]{article}
\begin{document}
\title{CAPACITANCE OF A DOUBLE-HETEROJUNCTION GaAs/AlGaAs
STRUCTURE SUBJECTED TO IN-PLANE MAGNETIC FIELDS:
RESULTS OF SELF-CONSISTENT CALCULATIONS}
\author{ T. Jungwirth,  L. Smr\v{c}ka \\
{ \it Institute of Physics, Acad. of Sci. of Czech Rep.,} \\
{\it Cukrovarnick\'{a} 10, 162 00 Praha  6, Czech Republic}}
\date{ (Received 23 July 1993)}
\maketitle
\vspace{1mm}
\noindent
{\bf Abstract.} The capacitance  of a  double-heterojunction structure
with a wide GaAs undoped layer embedded between two selectively
doped   AlGaAs  barriers   is  calculated   self-consistently  as
a function  of intensity  of  the  in-plane magnetic  field. With
increasing  field intensity  the capacitance  initially increases
and after  reaching a maximum decreases  toward a high  field limit
which  is  less  than  its  zero  field  value. This behaviour is
attributed to \lq breathing\rq{},  or charge redistribution, of
the 2D electron   gas  at   individual  heterojunctions   due  to
a combination  of the  confining potential  and the  magnetic field.

\vspace{10cm}

{\sc SUPERLATTICES and MICROSTRUCTURES}, in press
\thispagestyle{empty}
\newpage

\begin{center}
{\large \bf 1. Introduction}\\
\end{center}

\noindent
Recent technological developments in the domain of semiconductors
have  led  to  tremendously   growing  interest  in the (quasi)  two
dimensional electron  gas confined to  the $x -  y$ plane of  the
GaAs/AlGaAs interface by a narrow quantum well. Electrons possess
only two degrees of freedom along  the $x$ and $y$ direction, the
motion in the $z$ direction, perpendicularly to the interface, is
quantized.  A  rough  estimation  of  the  separation between the
lowest energy  levels is about $20$meV  in a \mbox{standard} GaAs/AlGaAs
heterostructure, the amplitude of the motion in the $z$ direction
is   around  $10$nm.   Thus  the   considered  systems   are  not
two-dimensional in  the strict sense, because  the wave functions
always have a  finite  spatial  extent  in  the  $z$ direction.

The magnetic  field $B$ applied  perpendicular to the  plane of the
2D gas further quantizes the energy spectrum into a set of Landau
levels separated by gaps  with the separation $\hbar\omega$ determined
by the cyclotron frequency $\omega  = |e|B/m$ proportional to the
magnetic  field.  The  quantization  of  the  kinetic  energy  of
electrons by the magnetic field is responsible for characteristic
magnetotransport phenomena, such as  Shubnikov-de Haas oscillations
\cite{1} and  the quantum Hall effect  \cite{2}. As the effective
mass $m$ of an electron is small in GaAs, the energy quantization
by experimentally available magnetic fields  is of the same order
of  magnitude  as  the  quantization  due  to  the  electrostatic
potential of a quantum well.

A magnetic field parallel to the interface does not introduce new
quantum  levels  but  strongly  influences  the  electron  energy
structure of a quasi 2D electron system both in  $k$-space and
in    real  space;  the  harmonic  magnetic  potential and the
confining  potential  of  a  quantum  well  are  combined into an
effective potential for the electron motion in the $z$ direction.
Obviously,  in  such a case  the  system  cannot  be considered as
a strictly two-dimensional one.

The most pronounced effect of a parallel magnetic field is a shift
of subband  energy dispersion curves \cite{3}.  The separation of
levels  and  the  density  of  states  in  branches of the energy
spectrum  increase  with  increasing  $B$.  As  a  result,  it is
possible to alter the individual  subband population to the point
where the highest subband is completely depopulated \cite{4}.

It was also shown that the  Fermi contours of the 2D electron gas
deviate  from  the  standard  circular  shape  under the combined
influence of  the confining potential  and the parallel  magnetic
field. In standard heterostructures with approximately triangular
potential  wells  the  Fermi  contours  acquire \lq pear-like\rq{}
shapes \cite{5}.

The theoretical investigations of the electronic structure of a 2D system
subjected to an in-plane magnetic field  were until now restricted to
simple  analytically  solvable  models  \cite{6,7,8},  or  to the
application of  perturbation  theory to more realistic quantum
mechanical  descriptions  of  systems  in  a  zero  magnetic field
\cite{3}. Only recently have fully self-consistent calculations of
the energy  subbands of standard  GaAs/AlGaAs heterostructures in
in-plane fields  been published \cite{9,10}.

In this paper we focus on  an interesting feature of the electronic
structure of  quasi 2D systems in  parallel magnetic fields which
did  not  attract  much  attention  until  now: the field induced
redistribution of the electron charge  density in a quantum well.
This  quantity is  routinely  calculated  in the  above mentioned
self-consistent calculations but rather difficult to measure. New
technique for  making separate connections  to the individual  2D
electron  layers \cite{11}  of double-quantum  wells makes  it at
present  experimentally  accessible  by  measuring  the  electric
capacitance  as  a  function   of  the  in-plane  magnetic  field
amplitude.

In  our theoretical  study of  the capacitance  we \mbox{shall} consider
a double-heterojunction  structure, where  the wide undoped GaAs
layer is  embedded between two selectively  doped AlGaAs barriers
(see  figure 1).  With an  appropriate choice  of the  GaAs
layer  width  $d$  this  configuration  produces two independent,
non-overlapping  2D  electron  systems  confined  to  GaAs/AlGaAs
interfaces  and   the  structure  resembles   a  parallel  plate
capacitor.

We assume  that our system is  symmetrical and has its  centre of
symmetry  at $z  =d/2$. Therefore,  in equi\-librium,   the
charge  distribution  also is  symmetrical  and  the  electro-chemical
potential  $\mu_L$ of  the left  electron layer  is equal  to the
electro-chemical potential $\mu_R$ of  the right layer. Note that
in this case  the electric field in the  middle of the structure,
at $z = d/2$, is equal to zero.

To   calculate   the   capacitance,   we   have   to   study  the
non-equilibrium state of a double-layer system with the left part
($z < d/2$) charged  $Q_L =  +Q$ and the right part ($z > d/2$)
charged    $Q_R  =  -Q$.  As  we  deal with two non-overlapping
electron  systems  (the  tunneling  between  them  is  completely
neglected) each of them can be assumed to be in a local thermodynamical
equilibrium,  characterized by  local electro-chemical potentials
$\mu_L$ and $\mu_R$, respectively. Then the  capacitance $C$ per  unit
area is given by

\begin{equation}
\label{-1}
C = \frac{|e|Q}{\Delta \mu}
\end{equation}

\noindent where $\Delta \mu = |\mu_L - \mu_R |$.

To determine  the capacitance defined  by the above  expression we
proceed in the following steps.

First,  basic  formulae  describing   the  electron  \mbox{subbands}  of
a quantum well subjected to the  in-plane magnetic fields will be
presented and we shall try to  conclude from their shape rather general
results independent  of a detailed  knowledge of the  form of the
confining potential. A simple analytically solvable model will be
used to illustrate this analysis.

Then the self-consistent calculation  of the equilibrium electronic
structure  of the  double-layered quantum  well will  be described  in
detail and, finally, the capacitance will be calculated, assuming
a difference  between electro-chemical  potentials of  individual
electron layers confined to GaAs/AlGaAs interfaces.\\

\begin{center}
{\large \bf 2. General formulae}\\
\end{center}

\noindent
We consider a system  of non-interacting electrons at each GaAs/AlGaAs
interface,  subjected  to  a  confining  electrostatic  potential
$V_{conf}(z)$  and  to an  in-plane  magnetic  field ${\bf B}\equiv
(0,B_y,0)$.  The main  difference between  the electrostatic  and
magnetic forces  is that the  electric field depends  only on the
$z$-coordinate  of an  electron while  the Lorentz  force is also
a function of the electron velocity components $v_x$ and $v_z$. Thus,
due  to the  presence of   the in-plane  magnetic field, time
reversal symmetry is broken, and the $x$ and $z$ components of the
electron motion couple.

Since the translational invariance in  the layer plane is preserved,
the  quantum  mechanical  description  of  the  system reduces to
a one-dimensional problem. In the Landau gauge, the
momentum components  $p_x$ and $p_y$ are
constants of  motion and the wavefunction  $\psi({\bf r})$ can be
factorized:

\begin{equation}
\psi({\bf r})=\frac{1}{\sqrt{S}}\,\exp(ik_xx+ik_yy)\,
\varphi_{i,k_x}(z).
\end{equation}

\noindent   Then  the   Sch\"{o}dinger  equation   for  the   \lq
out-of-plane\rq{} electron motion may be written as

$$
\left[-\frac   {\hbar^{2}}    {2m}   \frac   {\partial
^{2}}{\partial z^{2}}+ \frac{1}{2m}\left(\hbar k_{x}-|e| B_{y}
z\right)^{2}- e V_{conf}\left(z\right) \right]\varphi_{i,k_{x}}
\left(z\right)
$$
\begin{equation}
\label{0}
  =  \left[ E_{i}(k_x)  -\frac{\hbar^{2}k_{y}^{2}}{2m} \right]
 \varphi_{i,k_{x}}\left(z\right).
\end{equation}

\noindent  This one-dimensional  equation describes  the electron
motion  in  the  effective  electro-magnetic  potential $V_{eff}$
composed of  the harmonic magnetic  potential, corresponding to
the Lorentz force, and the confining potential $V_{conf}$

\begin{equation}
V_{eff} = \frac{m \omega^2}{2}\left(z -z_0\right)^{2} - e V_{conf}(z).
\end{equation}

\noindent The centre $z_0$ of  the magnetic part of the effective
potential is  related to the  wavevector component $k_x$  by $z_0
=\hbar  k_x/m\omega$.  This  feature  of  the effective potential
reflects  the  dependence  of  the  electron  motion  in  the $z$
direction on  the $x$ component of  the in-plane electron motion.
As  a  consequence, the eigenfunctions
$\varphi_{i,k_x}(z)$ become $k$-dependent.

When compared  with the zero  field solutions, the  $k$-dependent
eigenfunctions  are  modified  by   the  magnetic  field  in  two
different ways.

First,  the   centres  of  mass   $\langle  z\rangle_{i,k_x}$  of
$\varphi_{i,k_x}(z)$  are  shifted  from  their  original  positions
obtained for  $B = 0$. They  are related to the  $x$ component of
the in-plane velocity by

\begin{equation}
\label{1}
\langle z\rangle_{i,k_x} = \frac{\hbar k_{x}}{m\omega} -
\frac{\langle v_x\rangle_{i,k_x}}{\omega},
\end{equation}

\noindent  the  velocity   $\langle  v_x\rangle_{i,k_x}$  can  be
determined  from   the  shape  of   the  energy  spectrum   curve
$E_i(k_x)$ by

\begin{equation}
\langle v_x\rangle_{i,k_x} =
\frac{1}{\hbar}\frac{\partial E_{i}(k_{x}) }{\partial k_x} .
\end{equation}

Second, as the confining strengths  of the magnetic potential and
$V_{conf}$    are    added,    we    expect    that   the   width
$\Delta_{i,k_{x}}$ of states $\varphi_{i,k_{x}}(z)$ defined by

\begin{equation}
\Delta^{2}_{i,k_{x}} =
\langle z^{2}\rangle_{i,k_x} - \langle z\rangle_{i,k_x}^{2}
\end{equation}

\noindent will decrease with the  magnitude of the magnetic field
$B$. It can  be related to the energy  spectrum curve by equation
(\ref{1}) and the expression

\begin{equation}
\langle z^{2}\rangle_{i,k_x} = \frac{\hbar k_{x}}{m\omega}
\langle z\rangle_{i,k_x} -
\frac{1}{m \omega}\frac{\partial E_{i}(k_{x}) }{\partial \omega}.
\end{equation}

Both the shift of the  mass centre $\langle z\rangle_{i,k_x}$ and
the reduction   of   $\Delta_{i,k_{x}}$   will   contribute   to  the
redistribution  of the  electron charge  density by  the in-plane
magnetic field.

We  illustrate  the  changes  induced  by  the  magnetic field in
a simple analytically solvable model  of a parabolic quantum well
with the confining potential $V_{conf} = m\Omega^2z^2/2$. In this
case the centres  of mass are shifted from  the original positions
$\langle z\rangle_{0}=0$ linearly with respect to $k_x$

\begin{equation}
\label{2}
\langle  z\rangle_{i,k_x} = \frac{\omega}{\widetilde{\omega}^2}
\frac{\hbar k_x}{m}
\end{equation}

\noindent and the characteristic width of the wavefunction is

\begin{equation}
\label{3}
\Delta_{i,k_{x}} =\left(\frac{\hbar}{\widetilde{\omega}m}\right)^{1/2}.
\end{equation}

\noindent In the above  equations $\widetilde{\omega} = (\omega^2
+ \Omega^2)^{1/2}$.  Each  of the equations  (\ref{2}) and (\ref{3})
describes  one of  the above mentioned mechanisms  of the charge
redistribution  due to  the magnetic  field. It  follows from the
equation (\ref{2}) that  the centres of mass are  spread by the
magnetic field until the field reaches a critical value

\begin{equation}
\label{4}
B_c=\frac{m\Omega}{|e|}.
\end{equation}

\noindent Then, upon futher  increasing the magnetic  field, the
centres return to  the  point  $z=0$ in  the limit  of
infinite magnetic field. Due to the symmetry of the electrostatic
potential  the shifts  of the  centres corresponding  to $k_x$ and
$-k_x$ are  symmetrical with respect  to the point  $z=0$ and thus
the centre of mass of the whole system remains unchanged.

The  equation (\ref{3})  shows that  the characteristic  width of
wavefunctions decreases with  increasing magnetic field and in
the  limit of  infinite magnetic  field, the  wavefunctions become
$\delta$-functions.

One can expect this  type  of  behaviour of eigenfunctions  to be
preserved  for  any  \lq  reasonable\rq{}  confining potential of
a single quantum well. The centres  of mass, which were localized
around  a single  $\langle z\rangle_{i,k_x}$  at   zero magnetic
field, occupy a  finite range for $B \neq  0$. The occupied range
first increases  with increasing $B$, then  decreases and finally
is reduced to  a point for $B \rightarrow  \infty$. The positions
around which  $\langle z\rangle_{i,k_x}$ are  localized for $B  =
0$ and $B \rightarrow \infty$ need  not be the same in asymmetric
quantum wells.\\

\begin{center}
{\large \bf 3. Self-consistent confinement}\\
\end{center}

\noindent
The standard semi-empirical model  working quantitatively for the
lowest conduction states of  GaAs/AlGaAs heterostructures is used
to solve  the Schr\"{o}dinger equation  in the envelope  function
approximation \cite{12}.  The envelope function is  assumed to be
built from  host quantum states  belonging to a  single parabolic
band. The effect of  the effective mass mismatch is completely
neglected,  and  the  envelope  functions  of  GaAs and AlGaAs are
smoothly  matched at  the interface. Hence  the Schr\"{o}dinger equation
has a form given by (\ref{0}).

Since  we  are  interested  in  two  non-overlapping
double-layer electron
systems confined  to two identical GaAs\-AlGaAs  interfaces and since
the   whole    heterostructure   is   electro-neutral    and   in
thermodynamical   equilibrium,  each   of  the   systems  can   be
investigated separately.  We can restrict  ourselves to a  single
GaAs/AlGaAs heterojunction  and demand a  zero derivative of  the
confining potential $V_{conf}(z)$ at the point $z=d/2$.

The confining potential

\begin{equation}
V_{conf}(z)=V_b(z)+V_{s.c.}(z)
\end{equation}

\noindent is  a sum of (i)  the step function  $V_b(z)=V_b\,\theta(-z)$
corresponding to the conduction band discontinuity between AlGaAs
and GaAs (ii) a term  describing the interaction of an electron
with  ions  and (iii) the  electron-electron  interaction.  This term
should be calculated self-consistently and can be written as

\begin{equation}
V_{s.c.}(z)=V_H(z)+V_{xc}(z).
\end{equation}

\noindent The  Hartree term $V_H$ is  determined from the Poisson
equation

\begin{equation}
\label{5}
\frac{d^{2}V_{H}}{dz^{2}}=\frac{|e|\varrho(z)}{\varepsilon}
\end{equation}

\noindent and  for the exchange-correlation term  $V_{xc}$ we use
an  expression calculated  by Ruden  and D\"{o}hler  \cite{13} in
a density-functional formalism

\begin{equation}
V_{xc}\simeq-0.611\frac{e^2}{4\pi\varepsilon}
\left(\frac{3N_e(z)}{4\pi}\right)^{1/3}.
\end{equation}

\noindent  The conduction  band offset  $V_b$ and  the dielectric
constant   $\varepsilon$   enter   our   calculations   as  input
parameters. (We use mks units.)

For  modulation  doped  GaAs/AlGaAs  heterostructures,  the  total
charge density $\varrho(z)$ in  equation(\ref{5}) can be split
into   parts  corresponding   to  concentrations   of  electrons,
$N_e(z)$, their parent donors  in AlGaAs, $N_d^+(z)$, and ionized
residual acceptors in GaAs, $N_a^-(z)$:

\begin{equation}
\varrho(z)=e\left[N_e(z)-N_d^+(z)+N_a^-(z)\right].
\end{equation}

\noindent   The   usual   approximation   of   constant  impurity
concentrations is assumed.

In our calculation we consider a GaAs/AlGaAs heterostructure with
parameters  $N_d=2\times  10^{18}  \rm  cm^{-3}$, $N_a=10^{14}\rm
cm^{-3}$  and the  spacer thickness  $w=60 \rm  nm$ to obtain the
electron  system of  $N_e\approx 1.6\times  10^{11} \rm  cm^{-2}$
having one occupied  subband. We took the band  offset $V_b = 225
\rm meV$  and  the   dielectric  constant  $\varepsilon  =  12.9$
\cite{12}.

As  discussed  in  previous  sections  the  electronic structure of
a quasi  2D gas  is substantially  modified by  in-plane magnetic
fields both in   $k$-space and in real  \mbox{space}. Figure 2
shows    the    intervals     of    wavevectors    $k_x \in \langle
k_{x}^{min},\\ k_{x}^{max}\rangle$, determined by

\begin{equation}
E(k_{x}^{min})=E(k_{x}^{max})=\mu
\end{equation}

\noindent     and     corresponding     to     occupied    states
$\varphi_{k_x}\left(z\right)$,  together  with  the corresponding
centres  of  mass.  The  $\langle  z\rangle_{i,k_x}$  curves  are
plotted  for several  magnetic fields  and only  for the occupied
states $k_x\in\langle k_{x}^{min},k_{x}^{max}\rangle$.

The $k_x$ dependence of the centres  is non-linear and, due to the
asymmetry  of  the  self-consistent  confining  potential,  it is
asymmetric with respect to the origin of coordinates. Also the
critical  values of  the magnetic  field $B_c^R$  and $B_c^L$ for
electrons  belonging to  the right  and the left  part of  the energy
spectra  are different  ($B_c^R\approx 1.75\rm  T$, $B_c^L\gg8\rm
T$).

In  spite  of  many   quantitative  differences,  the  model  with
parabolic  confinement  gives  a  qualitative  explanation of the
numerical results. The electrons  with $k_x$ close to $k_x^{max}$
are  shifted into  the bulk  $\rm GaAs$  while the electrons with
$k_x$ close to $k_x^{min}$ are shifted toward the interface. It means
that these  two electron systems  are subjected either  to a weak
electric field $-dV_{conf}/dz$ in the bulk
$\rm  GaAs$ or a strong electric field $-dV_{conf}/dz$
at the interface. Describing  the electron systems separately, the
curve can be approximated by two  straight lines (see the case of
$B=4\rm T$)  corresponding to  weak  parabolic confinement with
$\Omega=\Omega^R$   and   strong   parabolic   confinement   with
$\Omega=\Omega^L$.  Since $\Omega^R\ll\Omega^L$  the critical values
of  the magnetic  field must  fulfill $B_c^R\ll  B_c^L$ as  follows
from  equation(\ref{4}).

With parabolic  confinement the  centre of  mass of  the whole
electron system is  independent of the magnetic field, due to the
symmetry of the potential. Figure 3 shows that in the $\rm
GaAs/AlGaAs$  single heterojunction  the situation  is different.
Here  the whole  electron system  is first  shifted into the $\rm
GaAs$  and  then  for  $B>B_c^R$   it  is  shifted  back  to  the
heterostructure interface.

 The distributions  of electrons at  $B=B_c^R$ and at  $B=8\rm T$
(the largest  magnetic field considered in  our calculations) are
compared in the  figure 4(a). We can see  the
remarkable charge  redistribution in  bulk  $\rm GaAs$. At the
interface, the effective electric field due to the confining potential
$V_{conf}(z)$ is so strong that  the redistribution of electrons by
 the magnetic  field is  negligibly small.
For this reason,  the
confining potential also  changes remarkably only deep in $\rm GaAs$ as
shown in the figure 4(b).\\

\begin{center}
{\large \bf 4. Capacitance}\\
\end{center}

\noindent
In    the   previous    section   an    equilibrium   state    of
a double-heterojunction  GaAs/AlGaAs structure  was investigated.
To find its capacitance we have to study a non-equilibrium system
assuming   a  difference   between  the   local  electro-chemical
potentials of the left 2D electron gas, $\mu_L$, and of the right
2D  electron gas,  $\mu_R$. Then  the confining  potential is  no
longer symmetrical  and a non-zero electric  field appears in the
middle of the sample at $z = d/2$ as shown in figure 5.

The  calculation  of  the
difference between electro-chemical potentials  can be split into
the following steps: (i) For a given  value of $Q$ the electric field
in  the  middle  of  the  sample,  i.e.  the  derivative  of  the
electrostatic potential at the point $z=d/2$, is determined by

\begin{equation}
\label{10}
\left.\frac{dV_{conf}}{dz}\right|_{z=\frac{d}{2}}=
\frac{|e|Q}{\varepsilon}.
\end{equation}

\noindent    Then   the    potentials   $V_{conf}^L(z<d/2)$   and
$V_{conf}^R(z>d/2)$ are  calculated separately with  the boundary
condition  (\ref{10}). To  be definite,  we choose  the otherwise
arbitrary origin of  the energy scale in such  a way that $\mu_L$
is \mbox{equal} to  zero and $\mu_R$ enters the  self-consistent loop as
a parameter to be determined. (ii) The electro-chemical potential
$\mu_R$ is obtained from the matching condition

\begin{equation}
V_{conf}^L\left(d/2\right)=
V_{conf}^R\left(d/2\right).
\end{equation}

\noindent In numerical practice we started with  $\mu_R^s= 0$,
similarly as in the case of $\mu_L$, and calculate the difference
between the resulting values $V_{conf}^L$ and $V_{conf}^R$ at $z =d/2$.
Then obviously we have

\begin{equation}
\Delta\mu = |V_{conf}^L\left(d/2\right)-
V_{conf}^R\left(d/2\right)|
\end{equation}

\noindent and the capacitance $C$ can be calculated from equation
(\ref{-1}).  We have  performed the  calculation for  a number of
concentrations of  the non-equilibrium charge $Q$  up to $0.5\%$ of
the equilibrium  concentration of electrons $N_e$,  for which the
value of
$|e\Delta\mu|$ reaches  $\approx 1 \rm  meV$. We have  found that
the capacitance is a constant and  does not vary with $Q$ in this
range of values.

The capacitance as  a function of the magnetic  field is shown in
figure 6. It reaches its maximum value at $B=B_c^R$ and
for   $B>B_c^R$  decreases   toward  a   limiting  value   $C=C_\infty$
for  $B \rightarrow  \infty$. The
relation of the shape of this curve to the magnetic field induced
charge  redistribution is  explained in   figure 7.
For magnetic fields $B <  B_c^R$ the electron systems are shifted
closer  to each  other and  thus the  potentials $V_{conf}^L$ and
$V_{conf}^R$   increase  at   the  point   $z=d/2$  (see   figure
7(a)). At the same time, since the electron concentration in the
right part  of the system is  larger than in the  left one, we can
expect  the  effect  of  the  charge redistribution  to be stronger on
$V_{conf}^R$ than  on $V_{conf}^L$. It means  that the difference
$V_{conf}^L-V_{conf}^R$  decreases  and  the  capacitance  of the
system increases. For $B>B_c^R$ the situation is just opposite, as
shown in the figure 7(b).

In previous sections  we mentioned that in the  limit of infinite
magnetic  field the  electron  systems  are compressed  to planes
close to the interfaces. It enables one to find the approximate limiting
value  of  the  capacitance
($C_\infty\approx 0.57\times\rm pF\mu m^{-2}$)
from  a  model  of  a parallel plate
capacitor with the distance between  plates equal to the distance
between the $\rm GaAs/AlGaAs$ interfaces.\\

\begin{center}
{\large \bf 5. Summary}\\
\end{center}

\noindent
The  capacitance of  bilayer two-dimensional  electron systems as
a function of  in-plane magnetic  field has been determined by
the   self-consistent calculation of the  electron structure
of a GaAs/AlGaAs double-heterojunction.

To gain better physical insight  into the problem, the qualitative
aspects of  the behaviour of  the system in  magnetic fields have
been illustrated using a simple model with parabolic confinement.

The  redistribution  of  the  charge  due  to  the
combined effect of the magnetic field
and  of the shape of the confining
potential was found responsible for the  results obtained.

We  have   also  found  that  the   self-consistent  approach  is
unavoidable in  calculation of the  difference between the  local
non-equilibrium  electro-chemical  potentials  of the two 2D systems
which define the capacitance of a bilayer system.\\

\noindent {\bf Acknowledgements-} We are grateful to Prof. Allan
H. MacDonald who turned our attention to this problem.

This work has been partly supported by ASCR Grant No. 11 059.


\newpage
\thispagestyle{empty}

\newpage
\begin{center}
{\large \bf Figure captions}\\
\end{center}

\noindent
{\bf Figure 1.} Schematic of the GaAs/AlGaAs double-heterojunction with
non-overlaping bilayer 2D electron systems subjected to the in-plane
magnetic field ${\bf B}$.
The electron systems have one occupied level $E_0<\mu$.
$V_0$ is the conduction band discontinuity, $w$ the spacer thickness
and $d$ the distance between interfaces.\\

\noindent
{\bf Figure 2.} Self-consistently calculated $k_x$
dependence of the centre
of mass of the wavefunction in a GaAs\-/AlGaAs
heterojunction (full lines).
The discrete values of the in-plane magnetic field are taken from $0$
to $8\rm T$. Intervals of $k_x$ corresponds to occupied states. Dashed
lines are related to models with parabolic confinement
$(\Omega^R\ll\Omega^L)$.\\

\noindent
{\bf Figure 3.} Self-consistently calculated in-plane
magnetic field dependence
of the centre of mass of the whole system. The maximum of the curve
corresponds to the critical value of the magnetic field
$B_c^R\approx 1.75\rm T$.\\

\noindent
{\bf Figure 4.} Self-consistent (a) charge distribution  and
(b) confining potential
 at $B=1.75\rm T$ (solid line) and $B=8\rm T$ (dashed line).\\

\noindent
{\bf Figure 5.} Schematic of the GaAs/AlGaAs double-heterojunction in a
non-equilibrium state with the left part charged  $+Q$ and the right
part $-Q$. The local thermodynamical equilibrium of the
bilayer electron
systems is characterized by the local
electro-chemical potentials $\mu_L$,
$\mu_R$.\\

\noindent
{\bf Figure 6.} Calculated in-plane magnetic field dependence
of the capacitance
of the bilayer 2D electron systems. The maximum of
the curve corresponds to
the critical value of the magnetic field $B_c^R$.\\

\noindent
{\bf Figure 7.} Schematic of the change of the left
and right part of the
confining potential and consequently of $\Delta\mu$ due
to the magnetic
fields (a) $B<B_c^R$  and (b) $B>B_c^R$. For $B<B_c^R$, $\Delta\mu$
decreases
with increasing the magnetic field, for $B>B_c^R$ $\Delta\mu$ increases
with increasing the magnetic field.
\end{document}